\documentclass[a4paper,11pt]{article}
\usepackage{jinstpub} 
\usepackage{silence}
\usepackage{caption}
\usepackage{subcaption}

\usepackage{hyperref}
\usepackage{academicons}
\usepackage{xcolor}
\usepackage{graphicx}
\usepackage{enumitem}
\usepackage[ruled,vlined]{algorithm2e}
\newcommand{\ms}{\text{ms}}     

\newcommand{\orcid}[1]{\href{https://orcid.org/#1}{\includegraphics[height=1em]{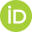}}}

\title{\boldmath $B$-jet Tagging Using a Hybrid Edge Convolution and Transformer Architecture }

\author{Diego F. Vasquez Plaza\orcid{0009-0007-3223-5815} and} 
\author{Vidya Manian \orcid{0000-0003-3834-8857}} 

\affiliation{Department of Electrical and Computer Engineering, University of Puerto Rico Mayaguez,\\ PR 00681-9000, USA}

\emailAdd{diego.vasquez@upr.edu}
\emailAdd{vidya.manian@upr.edu}

\abstract {Jet flavor tagging plays an important role in precise Standard Model measurement enabling the extraction of mass dependence in jet-quark interaction and quark-gluon plasma (QGP) interactions. They also enable inferring the nature of particles produced in high-energy particle collisions that contain heavy quarks. The classification of bottom jets is vital for exploring new Physics scenarios in the $\sqrt{s} = 13$~TeV proton-proton collisions. In this research, we present a hybrid deep learning architecture that integrates edge convolutions with transformer self-attention mechanisms, into one single architecture called the Edge Convolution Transformer (ECT) model for bottom-quark jet tagging. ECT processes track-level features (impact parameters, momentum, and their significances) alongside jet-level observables (vertex information and kinematics) to achieve state-of-the-art performance. The study utilizes the ATLAS simulation dataset \cite{data}. We demonstrate that 
ECT achieves 0.9333 AUC for $b$-jet versus combined charm and $light$ jet discrimination, 
surpassing ParticleNet (0.8904 AUC) and the pure transformer baseline (0.9216 AUC). 
The model maintains inference latency below 0.060~$\ms$ per jet on modern GPUs, 
meeting the stringent requirements for real-time event selection at the LHC. 
Our results demonstrate that hybrid architectures combining local and global features offer superior performance for challenging jet classification tasks. The proposed architecture achieves good results in $b$-jet tagging, 
particularly excelling in charm jet rejection (the most challenging task), while 
maintaining competitive $light$-jet discrimination comparable to pure transformer 
models.}


\arxivnumber{1234.56789} 

\begin{document}
\maketitle
\flushbottom

\section{Introduction}
\label{sec:intro}

Collimated showers of hadrons and leptons are called jets which are the dominant signatures 
of high energy quark and gluon production at the Large Hadron Collider (LHC). 
Identifying the flavor of the parton that initiated a jet (bottom, charm, or 
light quarks; gluons) is essential for precision Standard Model (SM) measurements 
and searches for physics beyond the SM. Heavy-flavor jets, 
particularly those originating from bottom-quarks ($b$-jets), play a central 
role in measurements of the Higgs boson ($H \to b\bar{b}$), top quark properties, 
and searches for supersymmetry.

The key discriminant for heavy-flavor tagging is the displaced secondary vertex 
arising from the finite proper lifetime of $b$- and $c$-hadrons 
($c\tau_b \approx 460~\mu\text{m}$, $c\tau_c \approx 150~\mu\text{m}$). Tracks originating 
from these decays exhibit large impact parameters relative to the primary 
interaction point. While distinguishing $b$-jets from $light$-flavor jets 
(u, d, s quarks, gluons) is relatively straightforward due to the absence of 
secondary vertices. However, separating $b$-jets from $c$-jets remains 
challenging due to their similar decay topologies. This $b$ vs $c$ discrimination 
is the limiting factor in many analyses, motivating continuous improvements in 
tagging algorithms.

Machine learning has transformed jet flavor tagging over the past decade, 
progressing from boosted decision trees operating on hand-crafted high-level 
features to deep neural networks processing low-level track and vertex 
information~\cite{Mondal:2024hf}. Recent state-of-the-art approaches employ 
either graph neural networks (e.g., ParticleNet~\cite{particleNet}) that 
aggregate information from spatially nearby particles, or transformer 
architectures~\cite{ParT} that capture global correlations via self-attention. 
Each paradigm has complementary strengths: graph convolutions excel at modeling 
local vertex topology, while transformers capture jet-wide patterns. In this work, we present the Edge Convolution Transformer (ECT), 
a hybrid architecture that integrates both mechanisms within a unified model. 
ECT employs EdgeConv blocks~\cite{particleNet} to extract local geometric 
features from $K$-Nearest Neighbor ($KNN$) graphs in $(\eta, \phi)$ space, followed 
by transformer self-attention layers to capture global particle correlations. 
A learned class token aggregates particle-level information, which is fused 
with jet-level vertex statistics for final classification. We demonstrate that 
this hybrid approach achieves superior performance compared to pure graph-based 
(ParticleNet) and pure attention-based (Particle Transformer) baselines, 
particularly for the challenging $b$ vs $c$ separation task, while maintaining 
inference latency suitable for real-time LHC trigger systems.

The main contributions of this research are:
\begin{itemize}[itemsep=0pt, topsep=0pt]
\item a novel hybrid architecture (ECT) combining EdgeConv and transformer 
      attention for jet flavor tagging;
\item comprehensive evaluation on ATLAS simulation data across three binary 
      classification tasks ($b$ vs $c$, $b$ vs $light$, $b$ vs $c$+$light$ jets);
\item performance comparison against ParticleNet and Particle Transformer 
      baselines; 
\item inference latency analysis confirming deployment feasibility for 
      high-level trigger systems; 
\item demonstration that EdgeConv blocks are essential for heavy-flavor separation, 
while transformer attention excels at capturing jet-wide patterns.
\end{itemize}

The remainder of this paper is organized as follows: Section 2 reviews related 
work in jet flavor tagging and transformer architectures. Section~3 describes 
the dataset, model architecture, and training methodology. Section~4 presents 
experimental results and comparative analysis. Section~5 concludes with 
limitations and future directions.

\section{Flavor Tagging Literature Review}
ML methods for jet flavor tagging make use of track properties and reconstructed secondary vertices as additional information both in ATLAS and CMS experiments.  A Scodellaro proceedings review summarizes Run 2 developments in heavy-flavor tagging such as track-based, vertex, soft-lepton, and boosted topology taggers and outlines preparations for the high-luminosity LHC era \cite{Scodellaro:2017btagATLASCMS}.
The study in \cite{Aaboud:2018xwy} presents the ATLAS Run~2 $b$-tagging algorithms, combining low level track and secondary vertex taggers into multivariate classifiers. It reports $b$-tagging efficiencies using 13~TeV $pp$ collision data corresponding to $80.5~\text{fb}^{-1}$, showing good agreement between data and simulation across a wide jet $p_T$ range. For real-time inference, the ATLAS collaboration developed a fast neural network based $b$-tagger, \textit{fastDIPS}, for deployment in the high-level trigger during LHC Run~3 \cite{ATLAS:2023btagFastDIPS}. This tagger enables early rejection of $light$ jets, reducing input rates for hadronic $b$-jet events by a factor of five with only a $\sim$2\% drop in signal efficiency. Another significant innovation is the Recurrent Neural Network (RNN)‑based $b$‑tagging algorithm (ATL‑PHYS‑PUB‑2017‑003) that sequences charged particle tracks in jets to exploit track–track correlations and improve $b$‑jet identification without relying on secondary vertex reconstruction \cite{ATL-PHYS-PUB-2017-003}. 

In the case of CMS experiments, the study in \cite{Sarkar:2024Run3UParT} reviews the heavy‑flavor jet tagging advancements including the UnifiedParticleTransformer (UParT) architecture and presents validated performance comparisons and scale factors derived from $13.6$~TeV collision data collected during 2022–2023. ParticleNet has been used for $b$-jet tagging in CMS High level trigger in 2022 and 2023 \cite{CMS:Run3PNetBtag}. Another CMS study \cite{CMS:2018P05011} presents Run 2 heavy‑flavor jet identification methods including track-based, secondary vertex, soft lepton taggers, boosted “double‑b” taggers, and c‑jet tagging, achieving ~68\% $b$-jet efficiency at 1\% $light$-jet misidentification and reducing uncertainty to a few percent across 30–1000 ~GeV jet \(p_T\). The review in \cite{Malara:2024jetsSubstructure} provides a comparative summary of state-of-the-art jet substructure and flavor‑tagging methods used by ATLAS and CMS including attention‑based transformers, adversarial training, and advanced calibration techniques highlighting their strengths and limitations in LHC Run 3 analyses.

The review of ML methods for heavy-flavor jet tagging by Mondal and Mastrolorenzo systematically categorises heavy‑flavor jet tagging methods at the LHC into three generations ranging from Boosted Decision Trees (BDTs) and dense neural networks to graph and transformer‑based architectures highlighting their evolution, detector input representations, and calibration strategies across Runs 1–3 \cite{Mondal:2024hf}. These methods have seen an evolution for Run 1, Run 2, and Run 3 CMS and ATLAS experiments at the LHC. JetVLAD, a set‑based heavy‑flavor jet tagger built on the NetVLAD architecture that aggregates particle-level descriptors into fixed-length vectors for classification is proposed in \cite{jetvlad}. It demonstrates strong performance on simulated heavy‑ion jet samples, achieving heavy‑flavor identification efficiencies above 80\% with background rejection factors up to several hundred. The authors in \cite{Stein:2023robustJetTag} demonstrate that adversarial training significantly enhances the robustness of jet-flavor tagging neural networks, preserving high classification accuracy while reducing vulnerability to input perturbations by probing and smoothing the loss surface geometry. Deep neural networks trained on raw track and vertex level information perform similarly to traditional $b$ taggers by directly leveraging high-dimensional detector data, showing that adding low-level features significantly improves jet-flavor classification \cite{Guest:2016jetFlavor}.

 For future colliders, the study in \cite{BuarqueFranzosi:2022fkh} presents a fast simulation framework based on Delphes with modules for tracking, time-of-flight, and cluster counting, and evaluates jet flavor tagging using the ParticleNet graph neural network tagger. The DeepJet model introduced in \cite{Bols:2020DeepJet} leverages deep neural networks including separate branches for charged, neutral, and secondary vertex inputs to process all jet constituents without pre-selection, significantly improving heavy-flavor and quark/gluon tagging performance over previous approaches. More recently, in \cite{Guvenli:2024RetNetBTAG} the authors introduce Retentive Networks (RetNet) for efficient $b$‑jet identification in simulated 13~TeV $pp$ collision data, achieving competitive performance with only 330k parameters and offering an effective alternative to models like DeepJet and Particle Transformer (ParT). 

Transformer-based models are increasingly popular in $b$-jet tagging.  In \cite{blekman2025taggingquarkjetflavours}, the authors present the DeepJetTransformer with scaled-dot product attention and heavy-flavor transformer block developed for the Future Circular Collider (FCC) experiment simulations. In \cite{Hammad:2024transformerJetTag}, the authors review state-of-the-art attention-based transformer architectures for heavy‑flavor jet tagging, demonstrating improved classification performance and model interpretability through physics-informed network modifications and analysis of the decision-making process. The ATLAS Collaboration introduces GN2, a groundbreaking transformer‑based jet flavor tagging algorithm that leverages end‑to‑end low-level track information and physics-informed auxiliary objectives: replacing traditional vertex-based taggers and delivering validated performance improvements in both simulation and    $\sqrt{s} = 13.6$~TeV Run 3 data analyses \cite{ATLAS:2025GN2}. 




\section{Bottom-Jet Tagging Methodology}
\label{sec:methodology}
This section describes the dataset, followed by data preprocessing and jet feature construction, computation of per-particle feature embedding, edge convolution operation, transformer self-attention, aggregation, followed by classification using a Feed-Forward Network (FFN). Further optimization is applied to training loops and evaluation metrics are presented.

\subsection{Notation}

Throughout this section, we use the following notation:
\begin{itemize}[itemsep=1pt]
\item $B$: Batch size
\item $P$: Maximum number of tracks per jet (40)
\item $d$: Embedding dimension (128)
\item $KNN$: Number of nearest neighbors for EdgeConv (16)
\item $h$: Number of attention heads (8)
\item $\mathbf{H} \in \mathbb{R}^{B \times P \times d}$: Particle embeddings
\item $\mathbf{M} \in \{0,1\}^{B \times P}$: Track validity mask
\end{itemize}

\subsection{Dataset and Preprocessing}
\label{subsec:dataset}
\subsubsection{ATLAS Simulation Dataset}
We utilize the publicly available ATLAS simulation dataset (Zenodo record 
4044628)~\cite{data} consisting of jets sampled from $pp \to t\bar{t}$ events 
at $\sqrt{s} = 14$~TeV. Events were generated using \textsc{Pythia8}~\cite{pythia8} 
and processed through \textsc{Delphes}~\cite{delphes} fast simulation framework 
configured to emulate the ATLAS detector~\cite{atlas}. This dataset was used for secondary vertex finding in \cite{Shlomi_2021} using the set-to-graph model.

The dataset contains three jet flavor classes:
\begin{itemize}[itemsep=2pt]
\item $b$-jets: Originating from bottom-quarks ($c\tau_b \approx 460~\mu$m)
\item $c$-jets: Originating from charm quarks ($c\tau_c \approx 150~\mu$m)
\item $light$ jets: Originating from up, down, strange quarks and gluons
\end{itemize}

\noindent The data is partitioned into training (62.2\%), validation (18.9\%), and test 
(18.9\%) sets, as detailed in Tables~\ref{tab:total_jets} and \ref{tab:total_tracks}.

\subsection{Feature Engineering}
\label{subsec:features}
This section presents the track-level and jet-level features and the preprocessing methods before training the models.

\subsubsection{Track-Level Features}

Each jet contains a variable number of charged particle tracks (up to $P_{\max}= 40$), zero-padded for batch processing. Seven features per track are extracted, chosen for their discriminative power in heavy-flavor identification:

\begin{enumerate}[itemsep=1pt]
\item $p_T$: Transverse momentum (log-transformed and z-normalized)
\item $d_0$: Transverse impact parameter
\item $z_0$: Longitudinal impact parameter  
\item $d_0/\sigma_{d_0}$: Transverse impact parameter significance
\item $z_0/\sigma_{z_0}$: Longitudinal impact parameter significance
\item $\text{IP}_{3D}$: 3D impact parameter
\item $\text{IP}_{3D}/\sigma_{\text{IP}_{3D}}$: 3D impact parameter significance
\end{enumerate}

The impact parameters characterize the spatial displacement of charged particle tracks relative to the primary interaction vertex (PV), exploiting the finite lifetime of heavy-flavor hadrons ($c\tau_b \approx 450~\mu$m, $c\tau_c \approx 150~\mu$m) to distinguish their decay products from prompt particles~\cite{ATLAS:2019btagPerf, ATL-PHYS-PUB-2016-012}. Following the ATLAS perigee track parameterization~\cite{ATLASTracking}:

\begin{itemize}[itemsep=2pt]
\item \textbf{Transverse impact parameter ($d_0$):} Defined as the distance of closest approach of the track helix to the primary vertex in the $r$-$\phi$ plane (transverse to the beam axis). Mathematically, $d_0$ represents the signed perpendicular distance 
from the PV to the track trajectory at the point of closest approach. Typical values range from $\mathcal{O}(10~\mu\text{m})$ for prompt tracks to $\mathcal{O}(1~\text{mm})$ for displaced tracks from $b$-hadron decays. The transverse resolution achieves $\sigma_{d_0} \approx 10$--$20~\mu$m for high-$p_T$ tracks in the ATLAS Inner Detector.

\item \textbf{Longitudinal impact parameter ($z_0$):} Defined as the $z$-coordinate difference between the primary vertex and the point of closest approach along the beam axis direction. In ATLAS analyses, the quantity $z_0 \sin\theta$ is often used to account for the track polar angle $\theta$, providing a more uniform resolution across pseudorapidity. The longitudinal component is particularly sensitive to pile-up contamination from nearby $pp$ interactions.

\item \textbf{3D impact parameter ($\text{IP}_{3\mathrm{D}}$):} Constructed as the quadrature combination of both components:

\begin{equation}
\text{IP}_{3\mathrm{D}} = \sqrt{d_0^2 + z_0^2}
\end{equation}

While $\text{IP}_{3\mathrm{D}}$ is indeed composed of $d_0$ and $z_0$, and these quantities are correlated, they provide complementary discriminative information. 
The transverse component ($d_0$) is more robust against pile-up effects, while the longitudinal component ($z_0$) captures additional vertex displacement information. The ATLAS IP3D algorithm exploits both parameters using two-dimensional probability density functions (PDFs) that explicitly model their correlation structure~\cite{ATL-PHYS-PUB-2016-012}, achieving superior performance compared to using either component alone.
\end{itemize}

The significance of each impact parameter is defined as the ratio of the 
measured value to its uncertainty: $d_0/\sigma_{d_0}$, $z_0/\sigma_{z_0}$, and 
$\text{IP}_{3\mathrm{D}}/\sigma_{\text{IP}_{3\mathrm{D}}}$, where $\sigma$ denotes the track-by-track measurement uncertainty propagated from the covariance matrix of the track fit. The 3D significance is computed as:

\begin{equation}
\frac{\text{IP}_{3\mathrm{D}}}{\sigma_{\text{IP}_{3\mathrm{D}}}} = \frac{\sqrt{d_0^2 + z_0^2}}{\sqrt{\sigma_{d_0}^2 + \sigma_{z_0}^2}}
\end{equation}

Tracks from $b$-hadron decays typically exhibit significance values $|d_0/\sigma_{d_0}|>3$ and $|\text{IP}_{3\mathrm{D}}/\sigma_{\text{IP}_{3\mathrm{D}}}|>5$, while prompt tracks from the primary vertex cluster near zero with widths determined by detector resolution. In our dataset, $b$-jets show mean $|d_0/\sigma_{d_0}|$ values approximately 2--3 times larger than $light$ jets, with pronounced high-significance tails extending beyond $10\sigma$ 
(Figure~\ref{fig:track}). These discriminative distributions motivate the inclusion of all three impact parameter types and their significances as independent features, despite their mathematical correlation, as each captures distinct aspects of the track displacement topology relevant for heavy-flavor identification.

Impact parameters and their significances are critical for distinguishing decay vertices: 
heavy-flavor jets exhibit displaced secondary vertices due to non-zero proper lifetimes 
of $b$- and $c$-hadrons, resulting in tracks with larger impact parameters compared to 
prompt tracks from $light$ jets.

\subsubsection{Jet-Level Features}

Eight global jet observables complement track-level information:

\begin{enumerate}[itemsep=1pt]
\item $p_T^{\text{jet}}$: Jet transverse momentum (log-transformed)
\item $\eta^{\text{jet}}$: Jet pseudorapidity
\item $\phi^{\text{jet}}$: Jet azimuthal angle
\item $M^{\text{jet}}$: Jet mass (log-transformed)
\item $N_{\text{trk}}$: Number of associated tracks
\item $N_{\text{vtx}}$: Number of reconstructed secondary vertices
\item $L_{3D}^{\text{max}}$: Maximum 3D displacement of secondary vertices
\item $N_{\text{trk}}^{\text{vtx,max}}$: Maximum tracks per secondary vertex
\end{enumerate}

\noindent The vertex-related features ($N_{\text{vtx}}$, $L_{3D}^{\text{max}}$, 
$N_{\text{trk}}^{\text{vtx,max}}$) encode crucial information about secondary 
vertex topology, particularly discriminative for $b$ vs $c$ separation given 
the different decay lengths.

\subsubsection{Normalization and Preprocessing}
The following are the normalization and preprocessing steps applied to the features.
\begin{itemize}[itemsep=2pt]
\item \textbf{Track features:} Z-score normalization per feature across training set
\item \textbf{Jet $p_T$ and mass:} Log-transformation followed by z-normalization
\item \textbf{Vertex features:} Min-max scaling to $[0,1]$ range
\item \textbf{Padding:} Jets with $<P_{\max}$ tracks are zero-padded
\item \textbf{Masking:} Boolean mask $\mathbf{M} \in \{0,1\}^{P_{\max}}$ indicates 
valid tracks (1) vs padding (0)
\end{itemize}


\begin{table}[htbp]
\centering
\caption{Total jet feature count}
\label{tab:total_jets}
\begin{tabular}{l|c|c}
\hline
Data     &Jet feature &Total quantity\\
\hline
Train 62.23\%      &$p_T^{\text{jet}}$, $\eta^{\text{jet}}$, $\phi^{\text{jet}}$, $M^{\text{jet}}$, $N_{\text{trk}}$, $N_{\text{vtx}}$,  &1,071,940\\

Validation 18.88\% &$L_{3D}^{\text{max}}$, $N_{\text{trk}}^{\text{vtx,max}}$ &325,289 \\

    Test 18.88\%       & &325,290\\
\hline
\end{tabular}
\end{table}

\begin{table}[htbp]
\centering
\caption{Total track feature count}
\label{tab:total_tracks}
\begin{tabular}{l|c|c}
\hline
Data     &Track feature &Total quantity\\
\hline
Train      & &7,624,594\\

Validation &$p_T^{\text{trk}}$, $d0$, $z0$, $d_0/\sigma_{d_0}$, $z_0/\sigma_{z_0}$, $ip3D$, $\text{IP}_{3D}/\sigma_{\text{IP}_{3D}}$ &2,312,445 \\

Test       & &2,311,995\\
\hline
\end{tabular}
\end{table}

\subsubsection{Classification Tasks}

We evaluate the ECT model on three binary classification tasks with increasing difficulty:

\begin{itemize}[itemsep=2pt]
\item {$b$ vs $light$:} Baseline task, exploiting large topological differences
\item {$b$ vs $c$+$light$:} Realistic scenario for LHC analyses, combined background
\item {$b$ vs $c$:} Most challenging, distinguishing similar heavy-flavor jets
\end{itemize}

\subsubsection{Data Distributions}

\begin{figure}[htbp]
\centering
\includegraphics[width=.45\textwidth]{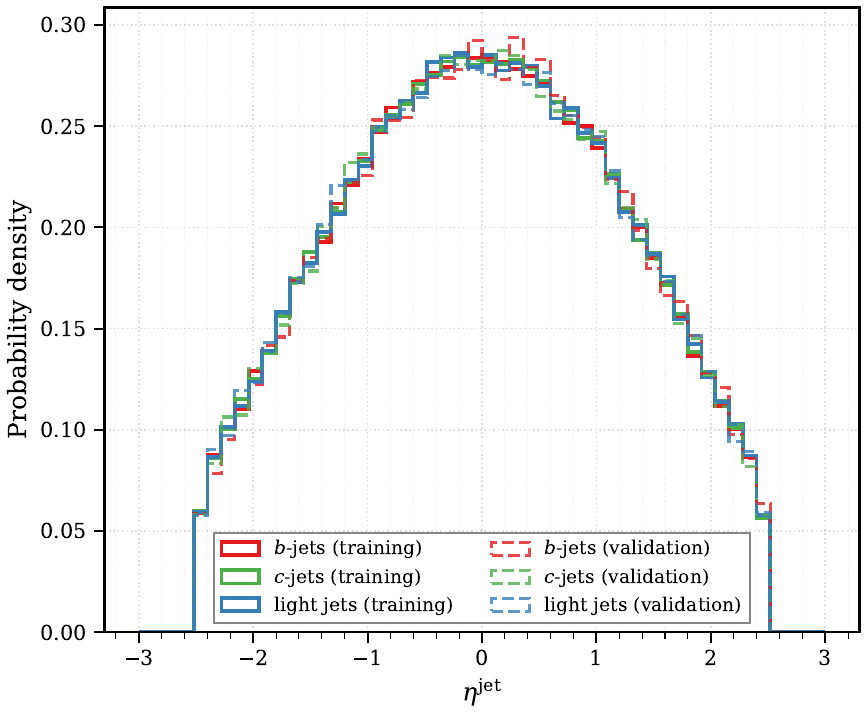}
\qquad
\includegraphics[width=.45\textwidth]{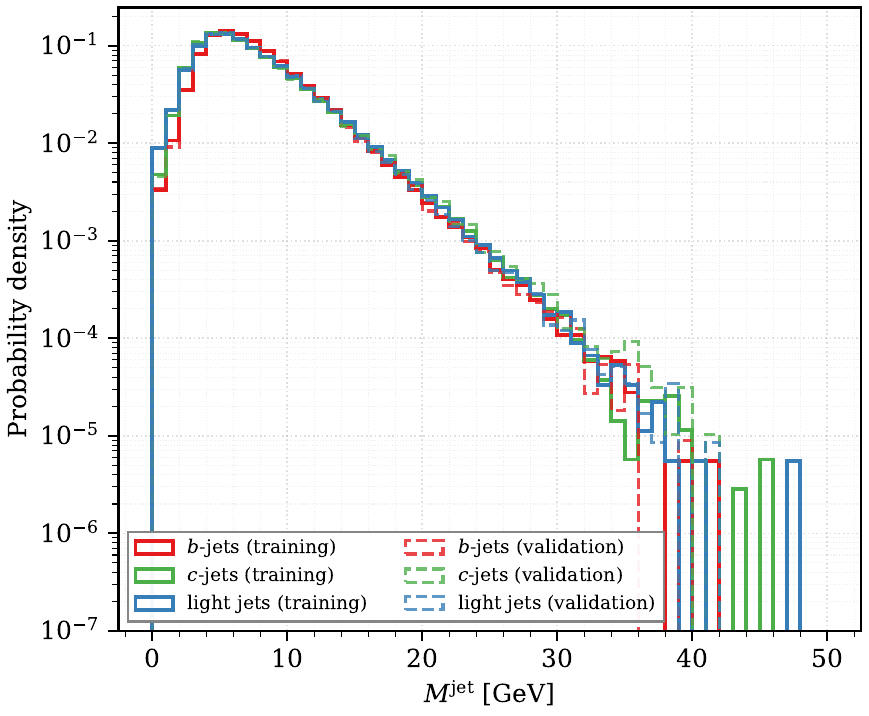}
\includegraphics[width=.45\textwidth]{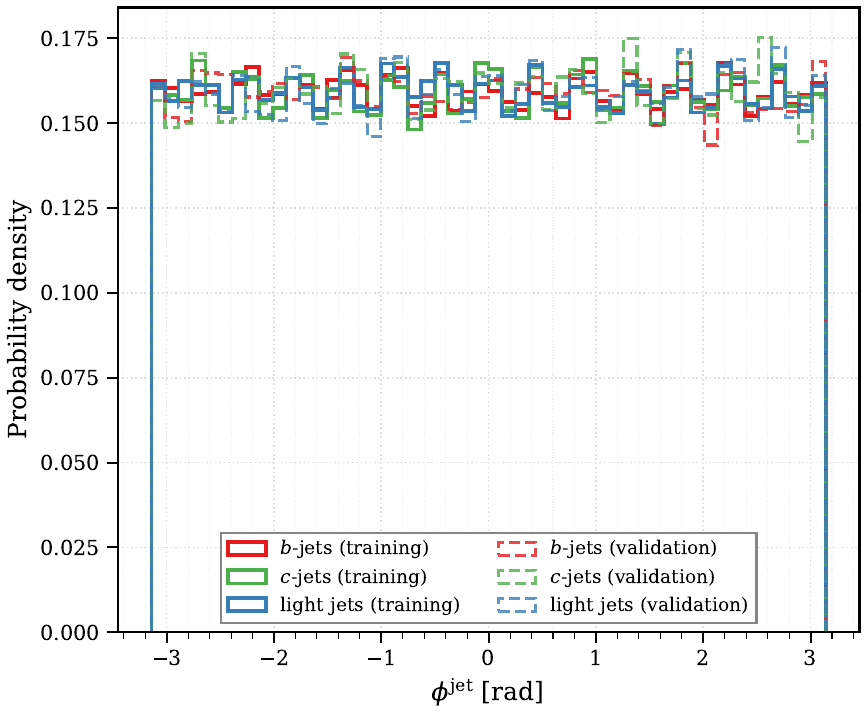}
\qquad
\includegraphics[width=.45\textwidth]{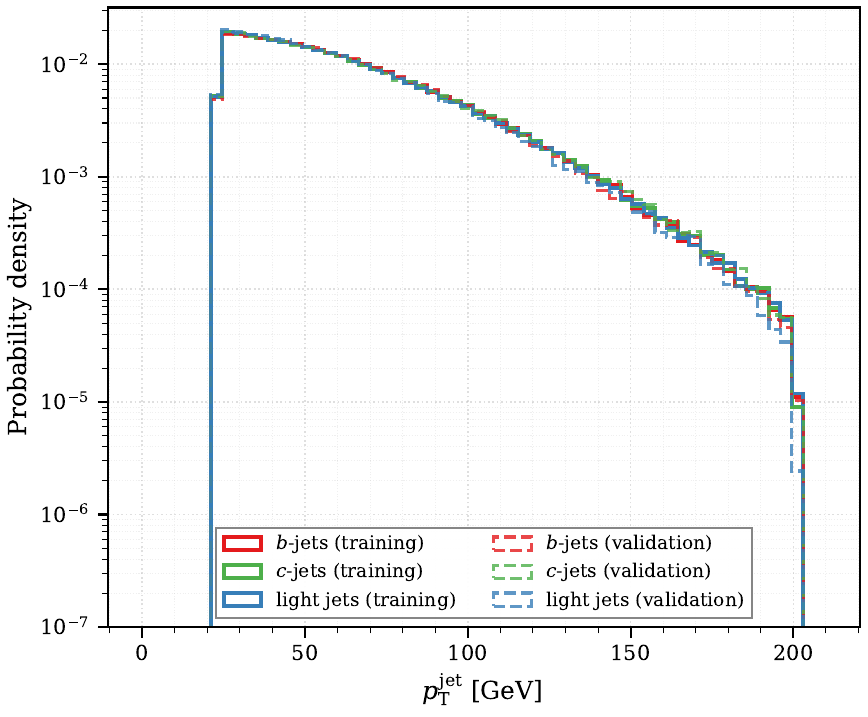}
\includegraphics[width=.45\textwidth]{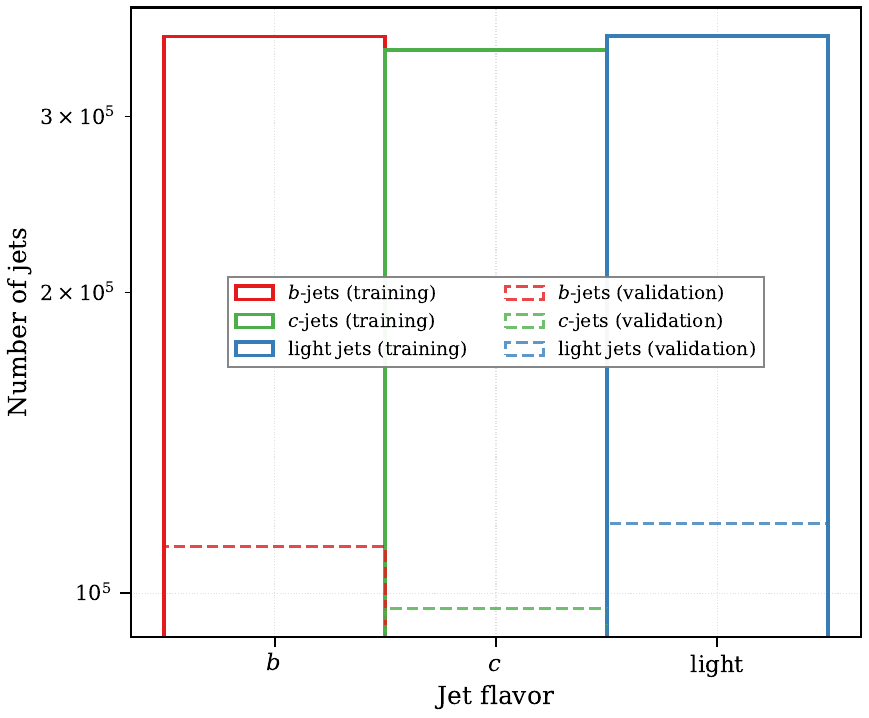}
\caption{
Distributions of jet-level features for $b$-jets (red), $c$-jets (green), 
and light jets (blue) in the ATLAS simulation dataset. 
\textbf{Top left:} Jet pseudorapidity ($\eta^{\mathrm{jet}}$). 
\textbf{Top right:} Jet invariant mass ($M^{\mathrm{jet}}$). 
\textbf{Middle left:} Jet azimuthal angle ($\phi^{\mathrm{jet}}$). 
\textbf{Middle right:} Jet transverse momentum ($p_{\mathrm{T}}^{\mathrm{jet}}$). 
\textbf{Bottom:} Jet flavor distribution. 
Solid lines represent the training set (62.2\% of data), while dashed lines 
show the validation set (18.9\%). All distributions are normalized to unit 
area for comparison. The broader $p_{\mathrm{T}}$ distribution of $b$-jets 
reflects the higher mass of bottom quarks ($m_b \approx 4.2$~GeV/$c^2$), 
while vertex multiplicity differences are evident in the flavor distribution.
\label{fig:jet}}
\end{figure}

\begin{figure}[htbp]
    \centering
    \includegraphics[width=.45\textwidth]{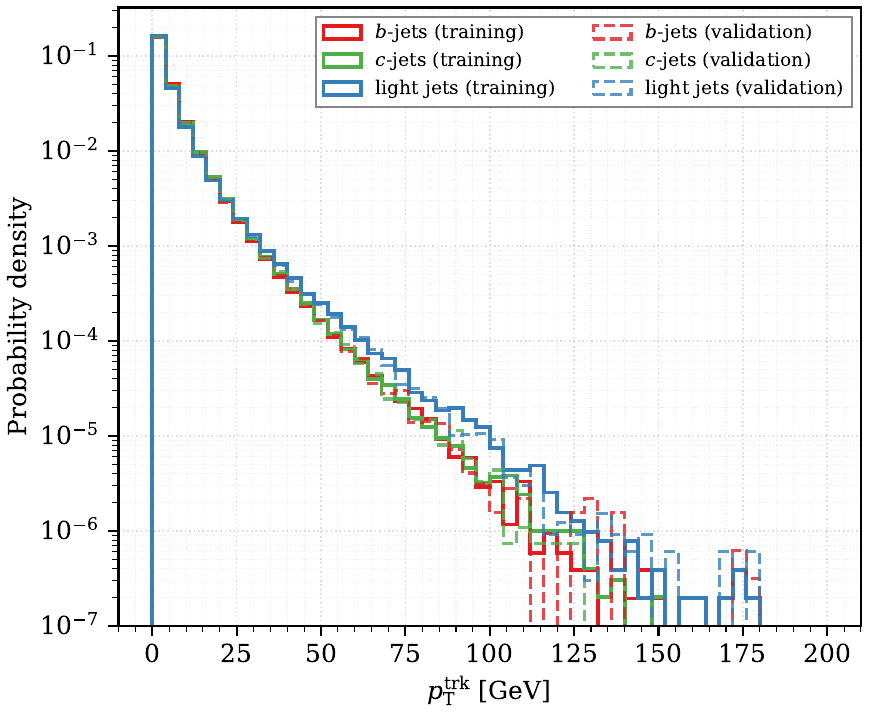}
\qquad
\includegraphics[width=.45\textwidth]{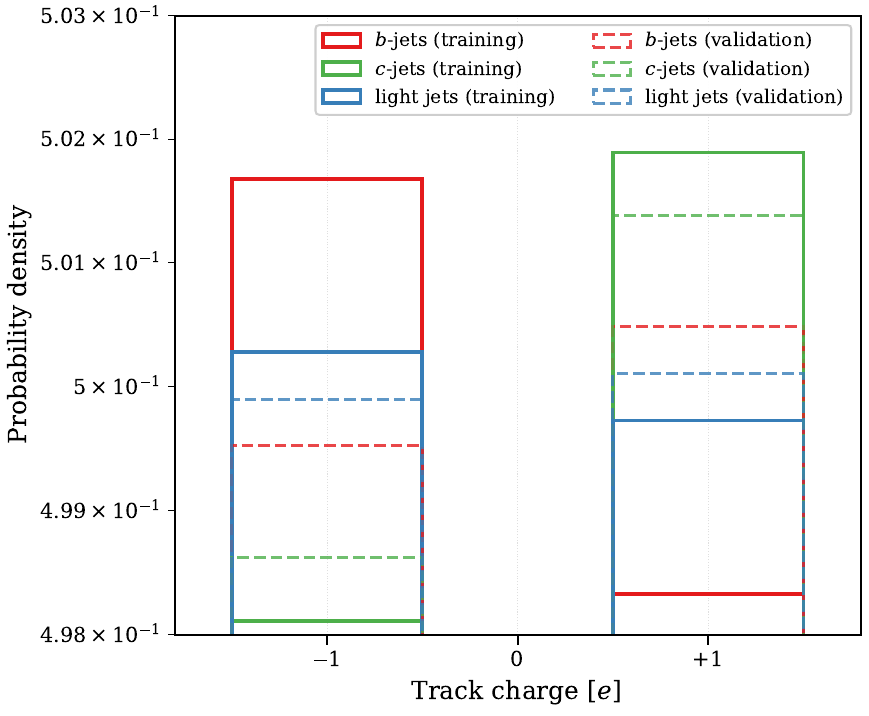}
\includegraphics[width=.45\textwidth]{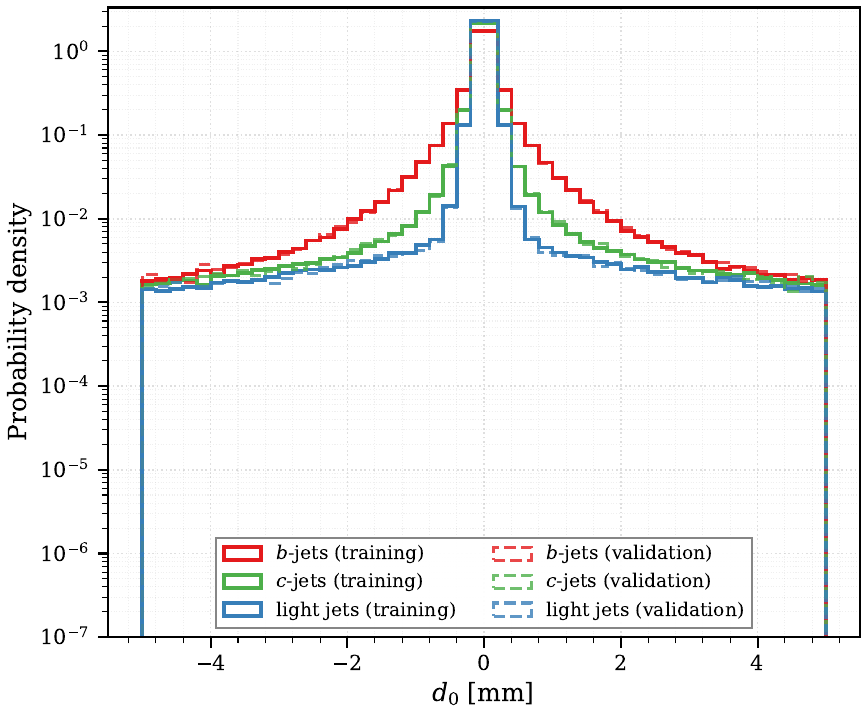}
\qquad
\includegraphics[width=.45\textwidth]{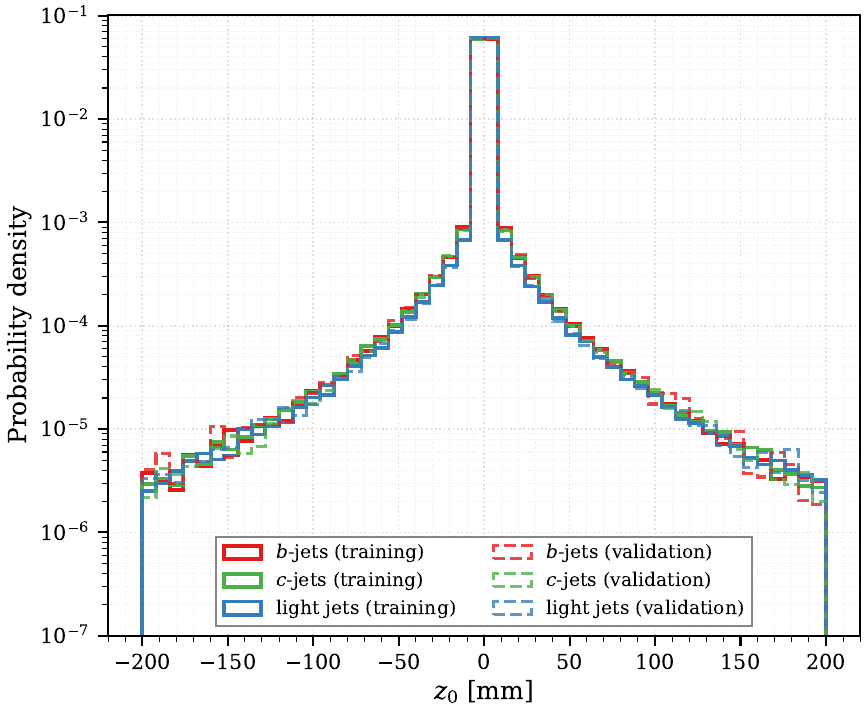}
\includegraphics[width=.45\textwidth]{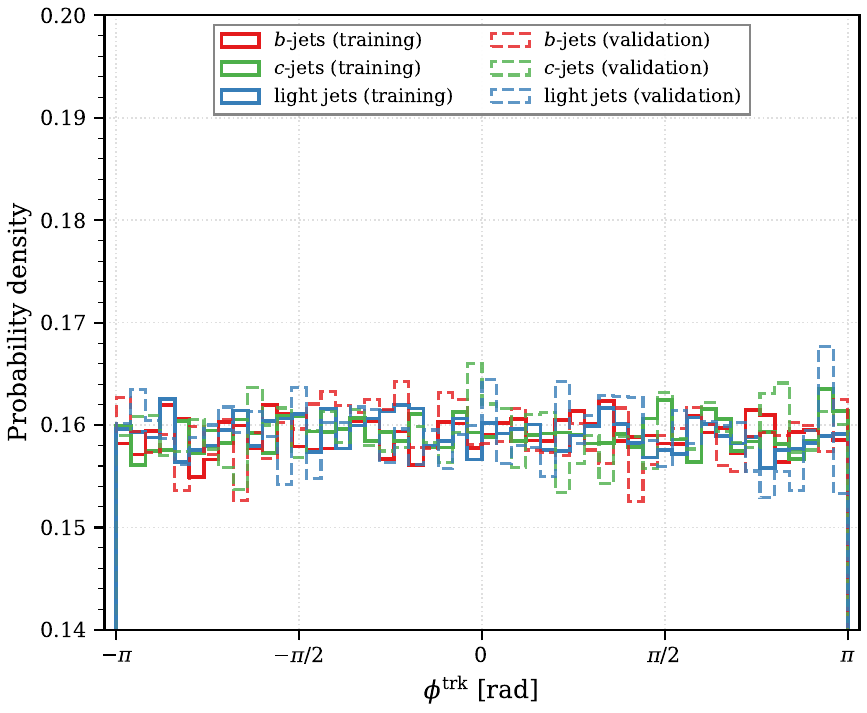}
\qquad
\includegraphics[width=.45\textwidth]{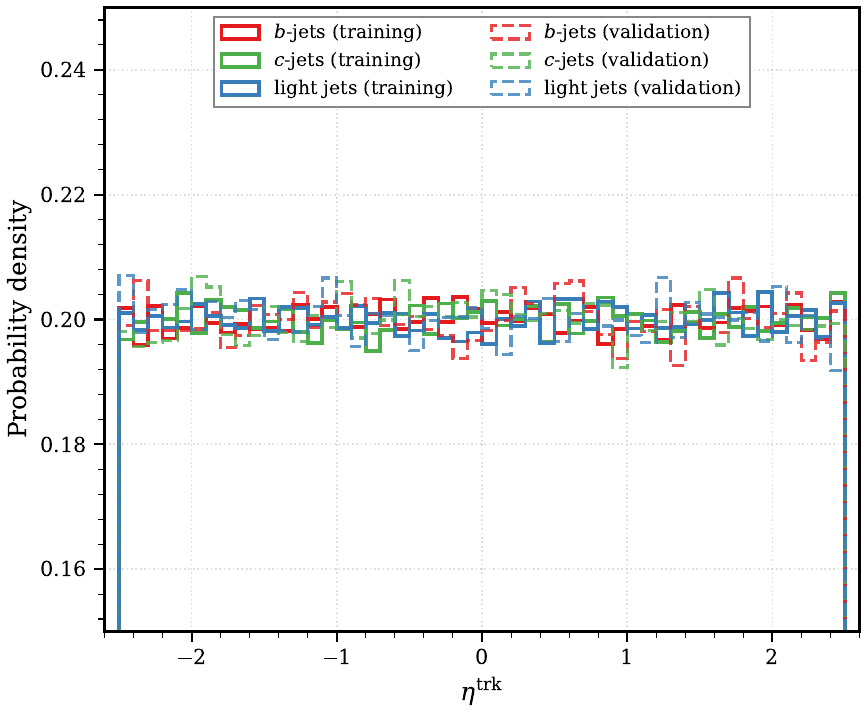}
    \caption{Distributions of track-level features for charged particles associated with $b$-jets (red), $c$-jets (green), and light jets (blue). \textbf{Top row:} Track transverse momentum ($p_{\mathrm{T}}^{\mathrm{trk}}$) showing $p_{\mathrm{T}}>1$~GeV/$c$ selection, and track electric charge distribution. \textbf{Second row:} Transverse ($d_0$) and longitudinal ($z_0$) impact parameters, demonstrating displaced vertices characteristic of heavy-flavor decays. The pronounced tails for $b$- and $c$-jets arise from secondary vertex displacements ($c\tau_b \approx 460~\mu$m, $c\tau_c \approx 150~\mu$m), while light jets peak sharply near zero, consistent with prompt tracks from the primary vertex. \textbf{Bottom row:} Track pseudorapidity ($\eta^{\mathrm{trk}}$) and azimuthal angle ($\phi^{\mathrm{trk}}$) showing angular distributions. where smaller uncertainties enable higher-significance discrimination of displaced vertices. Solid lines represent training data (7,624,594 tracks), dashed lines show validation data (2,312,445 tracks). All distributions are normalized to probability density. Heavy-flavor jets exhibit significantly broader impact parameter distributions, providing the primary discriminative power for $b$-jet identification}
    \label{fig:track}
\end{figure}

Figures~\ref{fig:jet} and \ref{fig:track} present the distributions of jet-level 
and track-level features across the three jet flavor classes. Several discriminative 
patterns are evident in Figure~\ref{fig:jet}: $b$-jets exhibit broader transverse 
momentum ($p_T$) distributions extending to higher values, reflecting the massive 
$b$-quark (mass $\approx$4.2~GeV); vertex multiplicity ($N_{\text{vtx}}$) is 
significantly higher for $b$-jets compared to charm and $light$ jets, consistent 
with the longer $b$-hadron lifetime enabling multiple displaced vertices to be 
reconstructed; jet invariant mass distributions show clear separation, with 
$b$-jets having systematically larger masses due to the heavier quark content.

Figure~\ref{fig:track} reveals the discriminative power of track-level features, 
particularly the impact parameter significance ($d_0/\sigma_{d_0}$, 
$z_0/\sigma_{z_0}$, $\text{IP}_{3D}/\sigma_{\text{IP}_{3D}}$). Heavy-flavor jets 
($b$ and $c$) exhibit pronounced tails extending to large significance values 
($>$10$\sigma$), corresponding to tracks originating from displaced secondary 
vertices. In contrast, $light$ jets show distributions tightly peaked near zero, 
consistent with prompt tracks pointing to the primary interaction vertex. The 
broader tails for $b$-jets compared to $c$-jets reflect the longer decay length 
($c\tau_b/c\tau_c \approx 3$), though substantial overlap makes pure cut-based 
separation challenging. These distributions motivate our choice of input features 
and demonstrate the rich discriminative information available at both track and 
jet levels for flavor classification.

\subsection{Architectural Overview}
\label{subsec:architecture_overview}

\begin{figure}[htbp]
\centering
\includegraphics[width=1\textwidth]{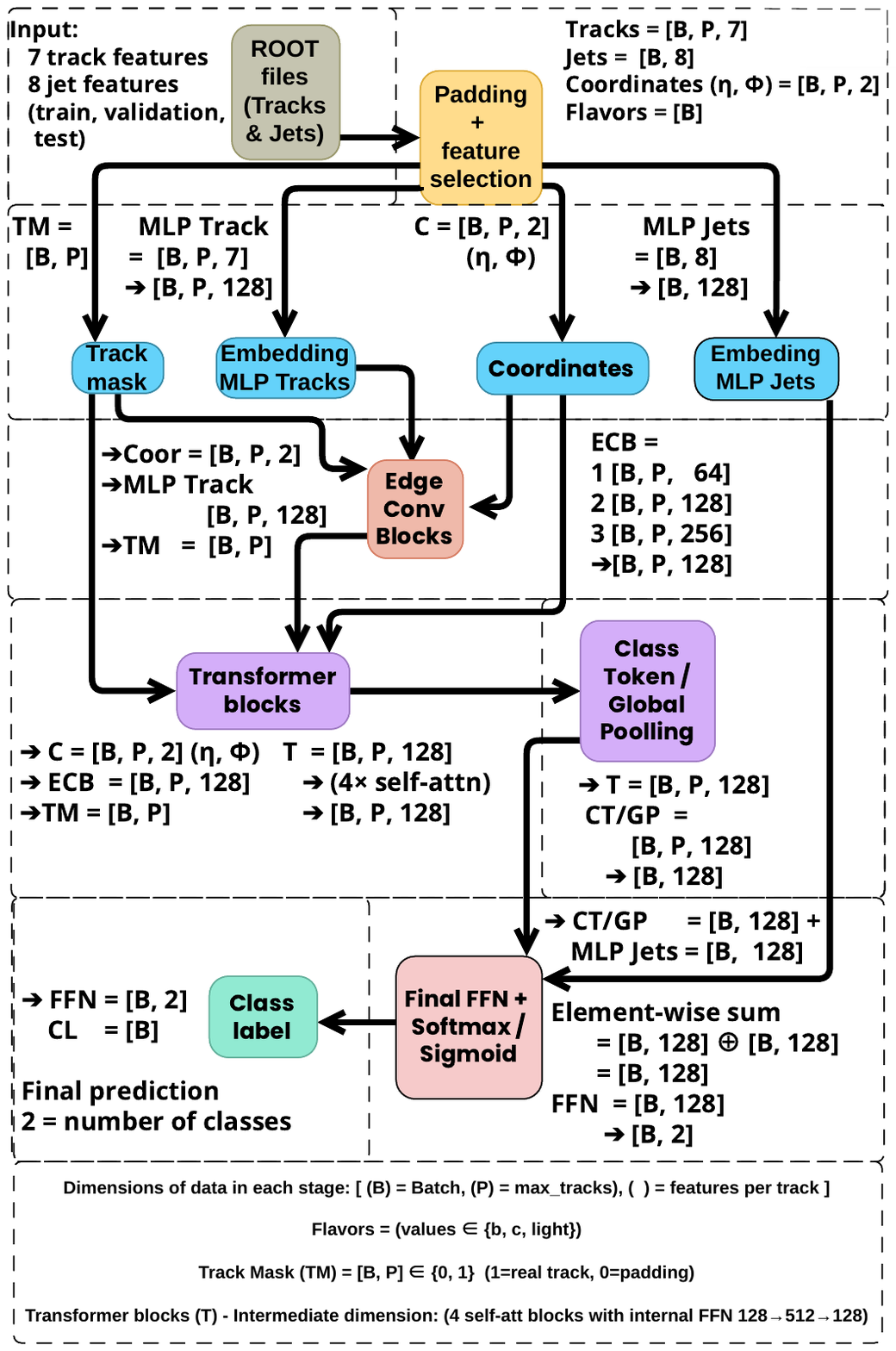}
\caption{Architecture of the Edge Convolution Transformer (ECT) for $b$-jet 
tagging, showing complete information flow and tensor dimensions at each stage. \\
}
\label{fig:arc}
\end{figure}

The ECT architecture processes jets through six sequential stages (Figure~\ref{fig:arc}):

\begin{enumerate}[itemsep=3pt]
\item \textbf{Input Processing:} In figure~\ref{fig:arc}, each jet contains up to $P = 40$ zero-padded tracks 
with 7 features per track (momentum and impact parameters), 8 jet-level global 
features (kinematics and vertex statistics), spatial coordinates $(\eta, \phi)$ 
for $KNN$ graph construction, and a boolean mask $\mathbf{M}$ distinguishing valid 
tracks from padding.
\item \textbf{Feature Embedding:} 
Track features are embedded from 7 to 128 dimensions via a three-layer Multi-Layer Perceptron (MLP) 
$[7 \to 128 \to 512 \to 128]$ with ReLU activations. Jet-level features are 
processed through a separate MLP $[8 \to 128 \to 128]$ with dropout ($p = 0.15$) 
for regularization.

\item \textbf{Local Feature Extraction:} 
Three EdgeConv blocks~\cite{particleNet} aggregate information from $KNN=16$ 
in $(\eta, \phi)$ space, capturing local jet substructure. 
The embedding dimension evolves through the blocks as $128 \to 64 \to 128 \to 256$, 
with residual connections and dimension-matching projections. A final linear layer 
projects back to $d = 128$.

\item \textbf{Global Interaction (Transformer):} 
Four self-attention 
layers with 8 heads each ($d_KNN = 16$ per head) capture long-range dependencies 
between particles. Each layer applies pre-LayerNorm, multi-head attention with 
scaled dot-product, and a position-wise FFN with hidden dimension 512 (4× embed 
dim). The track mask $\mathbf{M}$ ensures padded positions contribute zero attention weight.

\item \textbf{Jet-Level Aggregation:} 
A learned class token attends to all particle representations via two class-attention 
layers, producing a producing a permutation-invariant jet-level embedding 
$\mathbf{z}_{\text{cls}} \in \mathbb{R}^{128}$. This mechanism allows the model 
to learn which particles are most discriminative for flavor classification.

\item \textbf{Classification:} 
The class token representation $\mathbf{z}_{\text{cls}}$ 
is fused with the jet-level feature embedding $\mathbf{g}_{\text{jet}}$ via 
element-wise addition ($\oplus$), combining learned particle patterns with 
explicit vertex/kinematic information. A final linear classifier $[128 \to 2]$ 
with softmax activation produces class probabilities. Notation: Batch size $B$, 
maximum tracks per jet $P = 40$, embedding dimension $d = 128$. Total trainable 
parameters: 1,699,330 ($\approx$1.7M).
\end{enumerate}

\noindent The complete architecture contains 1.7M trainable parameters and is optimized end-to-end using the Adam optimizer with cross-entropy loss. Automatic Mixed Precision (AMP) training enables efficient utilization of GPU tensor cores.

\noindent \textbf{Design Rationale:} The hybrid design combines the complementary strengths 
of EdgeConv and transformers. EdgeConv captures local geometric relationships 
critical for vertex identification (e.g., clusters of displaced tracks), while 
transformer attention captures global patterns such as the overall track multiplicity 
and momentum flow characteristic of heavy-flavor decays. This combination proves 
particularly effective for the challenging $b$ vs $c$ discrimination task 
(Section~\ref{sec:results}).

\subsection{Training and Optimization}
The ECT model is trained using the Adam optimizer~\cite{adam} with  
learning rate of $\lambda_0 = 5 \times 10^{-4}$. We employ a constant learning rate schedule without warmup 
or decay, as preliminary experiments showed that simple schedules were sufficient 
for convergence. AMP training via 
torch.cuda.amp enables efficient utilization of GPU tensor cores, 
reducing training time by approximately 40\% with no loss in final model accuracy. A batch size of 1024 is used for the $b$ vs $c$+$light$ task due to 
its balanced class distribution, while batch size of 512 is employed for $b$ vs $c$ 
and $b$ vs $light$ to maintain stable training dynamics with more skewed class 
ratios. All models are trained for a maximum of 100 epochs with early stopping 
based on Area Under Curve (AUC). 

The hyperparameters for the ECT models were optimized empirically to balance training efficiency and performance. Table~\ref{tab:hyperparams} outlines the key hyperparameters used in the experiments for ECT along with those for ParticleNet and ParT models for comparison. The ECT architecture employs an embedding dimension of 128 with 4 multi-head attention heads. 
Four particle-level attention blocks and two class-level attention blocks process track and jet features. 
EdgeConv layers use $KNN=16$, with progressively deeper MLPs of sizes (64,64,64), (128, 128, 128), and (256, 256, 256). ReLU activations are used in the MLP embedding. The workflow for the ECT architecture is given in Algorithm \ref{alg:ect}.

\subsection{Evaluation Metrics} The evaluation metrics on the validation and testing set are classification accuracy, F1-score, Receiver Operating Characteristic (ROC), and AUC. If the AUC does not improve for 25 epochs, the training is stopped early, and the best model is restored. This model is used for the evaluation of the test set. The test set is evaluated by loading the best model checkpoint and computing the metrics of: accuracy, F1-score, and AUC. The ROC and confusion matrices are printed to visualize the performance of the ECT model.

\begin{table}[htbp]
\caption{Model hyperparameters and architectural configuration. ECT combines 
EdgeConv blocks (from ParticleNet) with transformer self-attention and 
class-attention mechanisms. All models trained on NVIDIA RTX A5000 GPU.}
\label{tab:hyperparams}
\centering
\small
\begin{tabular}{|l|c|c|c|}
\hline
\textbf{Parameter} & \textbf{ECT} & \textbf{ParticleNet} & \textbf{ParT} \\ \hline
\multicolumn{4}{|l|}{\textit{Architecture}} \\ \hline
Embedding dimension & 128 & 128 & 128 \\
Attention heads & 8 & -- & 8 \\
Self-attention blocks & 4 & -- & 8 \\
Class-attention blocks & 2 & -- & 2 \\
EdgeConv blocks & 3 & 3 & -- \\
EdgeConv $KNN$ & 16 & 16 & -- \\
EdgeConv channels & (64,64,64) & (64,64,64) & -- \\
 & (128,128,128) & (128,128,128) & -- \\
 & (256,256,256) & (256,256,256) & -- \\
Global MLP hidden dim & 128 & -- & 128 \\
Global dropout & 0.15 & -- & 0.15 \\
\textbf{Total parameters} & \textbf{1.7M} & \textbf{1.7M} & \textbf{1.7M} \\
\hline
\multicolumn{4}{|l|}{\textit{Training Configuration}} \\ \hline
Learning rate & $5 \times 10^{-4}$ & $1 \times 10^{-3}$ & $1 \times 10^{-3}$ \\
Batch size & 1024 / 512$^{\dagger}$ & 1024 & 1024 \\
Optimizer & Adam & Adam & Ranger \\
Max epochs & 100 & 100 & 100 \\
Early stopping patience &25 & 25 & 25 \\
LR scheduler & -- & -- & CosineAnnealing \\
Activation (embed MLP) & ReLU & ReLU & GELU \\
Mixed precision (AMP) & Yes & Yes & Yes \\
\hline
\multicolumn{4}{|l|}{\textit{Computational}} \\ \hline
GPU & \multicolumn{3}{c|}{NVIDIA RTX A5000 (24GB)} \\
Framework & \multicolumn{3}{c|}{PyTorch 2.0 + CUDA 11.8} \\
Training time ($b$ vs. $c$+$light$) & 2.2 hours & 4.5 hours & 1.5 hours \\
\hline
\end{tabular}

$^{\dagger}$ 1024 for $b$ vs. $c$+$light$, 512 for $b$ vs. $c$ and $b$ vs. $light$ tasks
\end{table}

\begin{algorithm}[t]
\caption{ECT: Jet Tagging Pipeline}\label{alg:ect}
\DontPrintSemicolon
\textbf{Part 1: Data Preprocessing}\;
\KwIn{Raw ROOT files with jet and track data}
\ForEach{jet event in dataset}{
  Extract track features $\{\mathrm{trk\_pt},
  \mathrm{trk\_d0}, \mathrm{trk\_z0}, \mathrm{trk\_d0sig}, \mathrm{trk\_z0sig}, \mathrm{ip3D}, \mathrm{ip3D\_signal}\}$\;
  Extract track coordinates $(\mathrm{trk\_eta}, \mathrm{trk\_phi}) $(for $KNN$ graph construction)\;
  {Extract jet features $\{\mathrm{jet\_pt}, \mathrm{jet\_eta}, 
  \mathrm{jet\_phi}, \mathrm{jet\_M}, \mathrm{n\_trks}, \mathrm{n\_vertex}, \mathrm{vertex\_L3D}, \mathrm{vertex\_ntracks}\}$}
  Pad tracks to fixed length $P_{\max} = 40$, construct mask\;
  Assign jet label according to classification mode (e.g., $b$ vs $c$)\;
}

\vspace{0.5em}
\noindent \textbf{Part 2: Training the Model}\\
\textbf{Require:} Preprocessed dataset $\mathcal{D}$, model $f_\theta$, loss function $\mathcal{L}$, optimizer\\
\ForEach{training epoch}{
  \ForEach{mini-batch $(\mathbf{X}, \mathbf{C}, \mathbf{M}, \mathbf{y})$ from $\mathcal{D}$}{
    Embed track features via MLP: $\mathbf{H} \gets \mathrm{MLP}(\mathbf{X})$\\
    \ForEach{EdgeConv block}{
      Update $\mathbf{H} \gets \mathrm{EdgeConv}(\mathbf{H}, \mathbf{C}, \mathbf{M})$
    }
    \ForEach{Transformer block}{
      $\mathbf{H} \gets \mathrm{SelfAttn}(\mathbf{H}, \mathbf{C}, \mathbf{M})$
    }
    Aggregate with class token or pooling: $\mathbf{z} \gets \mathrm{Aggregate}(\mathbf{H})$\\
    Predict logits: $\hat{\mathbf{y}} \gets \mathrm{FFN}(\mathbf{z})$\\
    Compute loss: $\ell \gets \mathcal{L}(\hat{\mathbf{y}}, \mathbf{y})$\\
    Update $\theta$ using optimizer and gradient of $\ell$
  }
  Validate on held-out data; apply early stopping if needed
}

\vspace{0.5em}
\noindent \textbf{Part 3: Evaluation and Inference}\\
\textbf{Require:} Trained model $f_\theta$\\
\ForEach{test jet event}{
  Preprocess features and coordinates as in Part 1\\
  Predict jet class: $\hat{y} = \arg\max\, f_\theta(\mathbf{X}, \mathbf{C}, \mathbf{M})$\\
  Collect predictions for metrics (accuracy, AUC, F1, etc.)
}
\end{algorithm}

\section{Bottom-Jet Tagging Results and Discussion}
\label{sec:results}

\begin{figure}[htbp]
\centering
\includegraphics[width=1\textwidth]{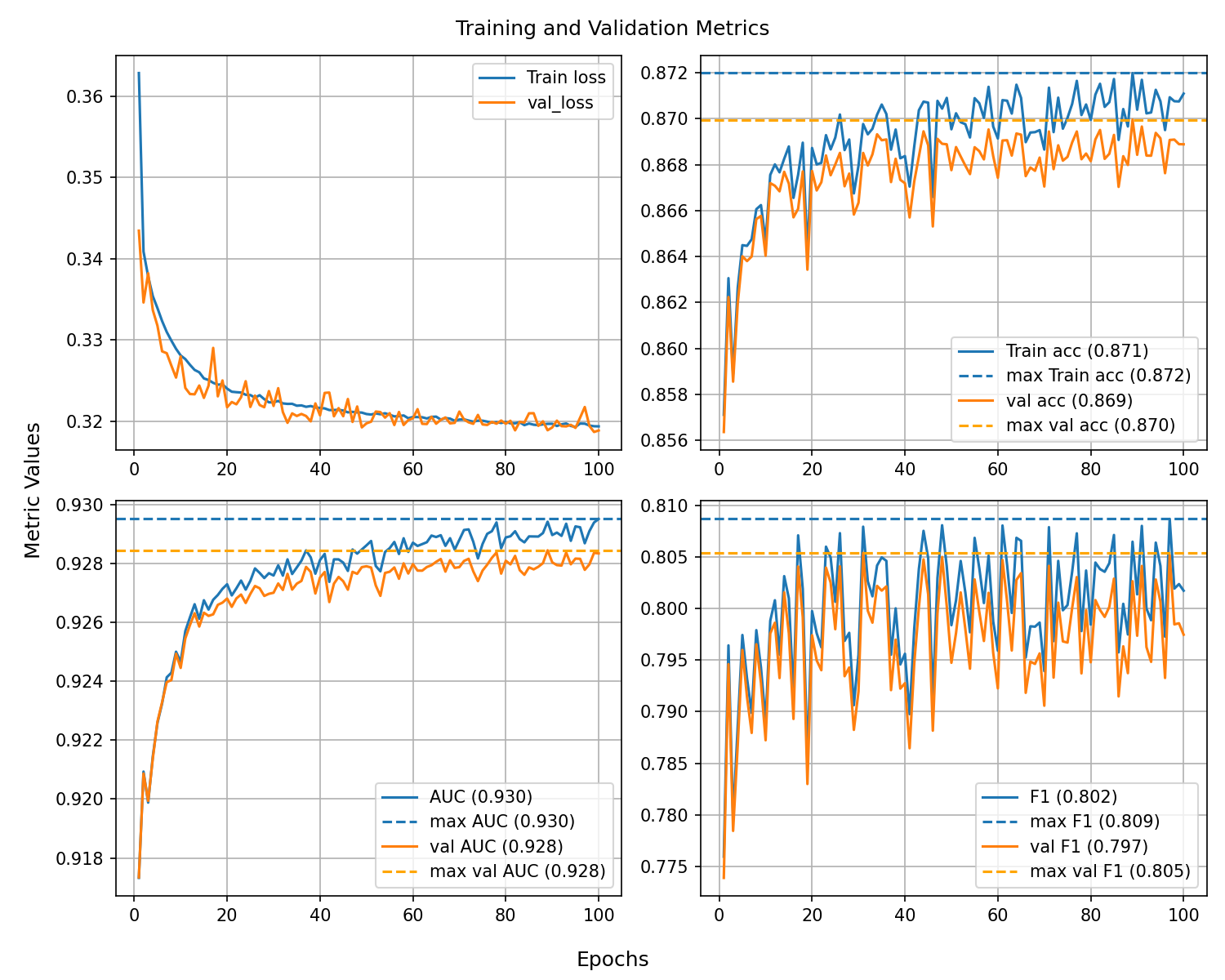}
\caption{Training (Train) and validation (Val) metrics for the ECT model on the $b$ vs $c$+$light$ 
classification task over 100 epochs. \textbf{Top left:} Cross-entropy loss 
converges smoothly from 0.37 to 0.31 over the first 50 epochs, then stabilizes, 
indicating effective optimization without oscillations. \textbf{Top right:} 
Training (blue) and validation (orange) accuracy (acc) reach $\sim$87\% with minimal 
gap (< 0.2\%), demonstrating that the model generalizes well without overfitting. 
\textbf{Bottom left:} Area Under Curve (AUC) improves steadily from 0.92 to 0.93, 
plateauing around epoch 89 where early stopping was triggered (patience = 25). The 
small train-validation AUC gap confirms robust generalization. \textbf{Bottom right:} 
F1-score exhibits higher variance during early training due to threshold 
sensitivity, then stabilizes at 0.81, confirming balanced precision-recall 
trade-off. Dashed horizontal lines mark the maximum values achieved: training 
accuracy 87.2\%, validation accuracy 87.0\%, training AUC 0.930, validation AUC 
0.928. These curves demonstrate stable convergence and effective regularization.
\label{fig:metrics}}
\end{figure}

\begin{figure}[htbp]
\centering
\includegraphics[width=0.9\textwidth]{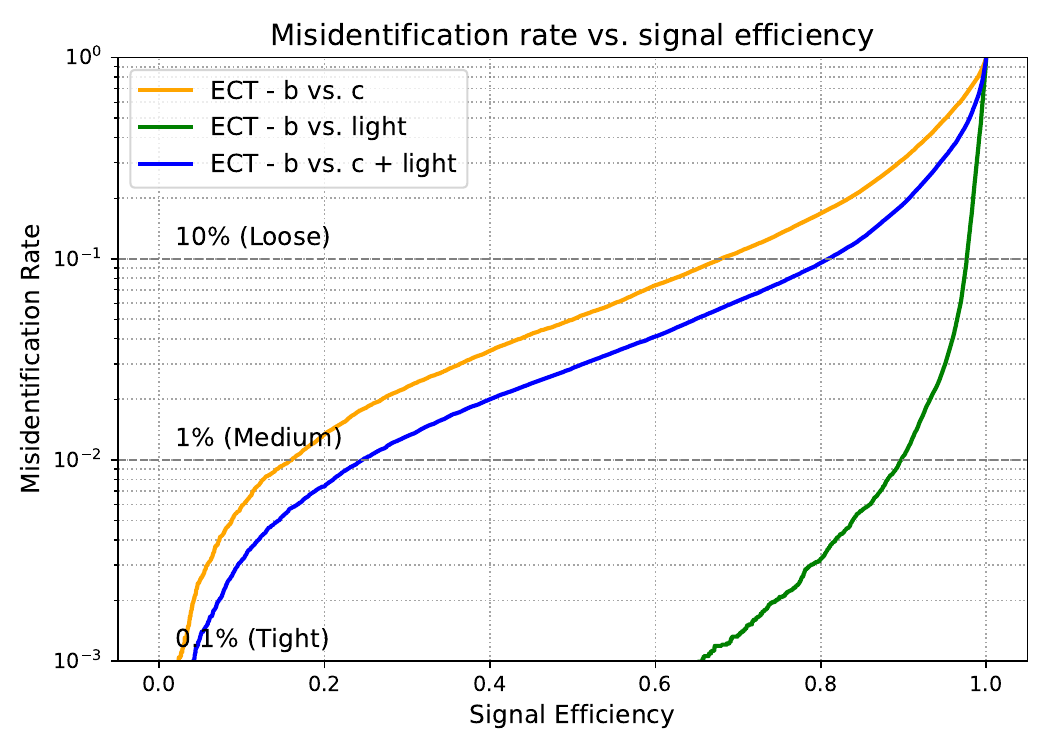}
\caption{Misidentification rate versus signal efficiency for the ECT model 
across three $b$-jet classification tasks: $b$ vs.\ $c$ (orange), 
$b$ vs.\ $light$ (green), and $b$ vs.\ $c$+$light$ (blue). 
Horizontal dashed lines indicate standard working points used in ATLAS analyses: 
Loose (10\% mistag rate), Medium (1\%), and Tight (0.1\%). 
The $b$ vs.\ $light$ discrimination achieves the best performance, 
reaching signal efficiencies above 65\% even at the Tight working point, 
reflecting the distinct signatures of $light$ jets compared to heavy-flavor jets.
The more challenging $b$ vs.\ $c$ task, where both jet types contain 
displaced vertices from heavy-hadron decays, shows reduced but still competitive 
performance. ECT achieves inference throughput of approximately 17,380 jets/s 
on a single NVIDIA RTX A5000 GPU, corresponding to less than 60~$\mu$s per jet, 
well within LHC High-Level Trigger latency requirements.
\label{fig:ECT}}
\end{figure}

\begin{figure}[htbp]
\centering
\includegraphics[width=0.9\textwidth]{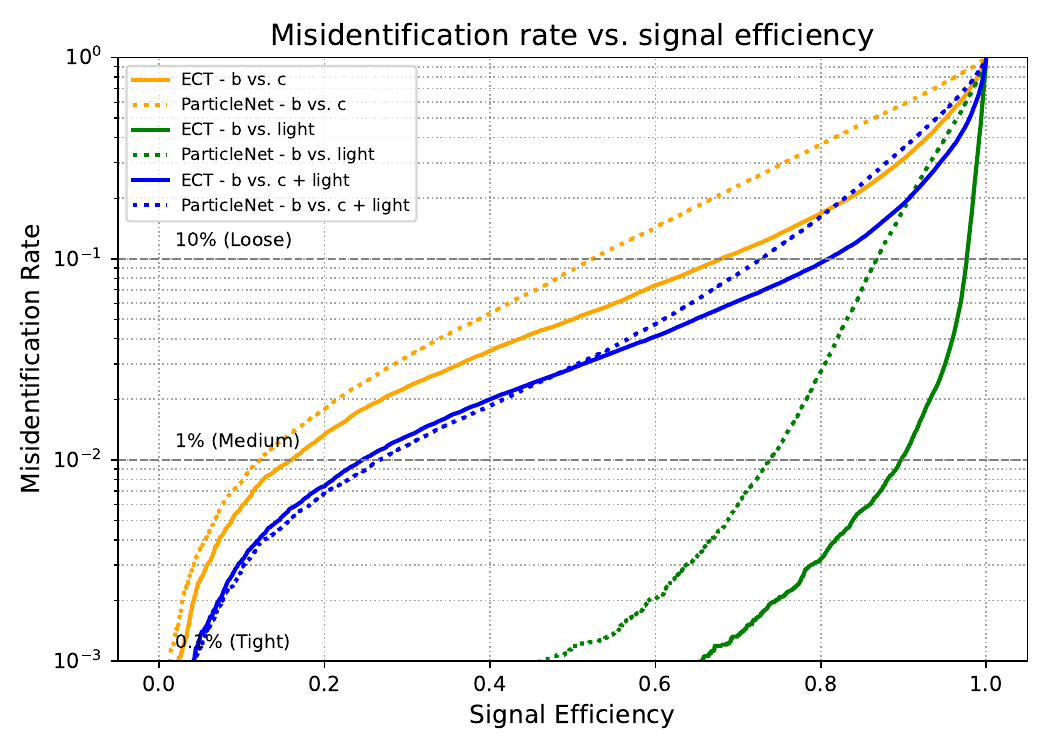}
\caption{Comparison of misidentification rate versus signal efficiency 
between ECT (solid lines) and ParticleNet (dotted lines) for $b$-jet tagging.
Three classification scenarios are evaluated: $b$ vs.\ $c$ (orange), 
$b$ vs.\ $light$ (green), and $b$ vs.\ $c$+$light$ (blue).
ECT consistently outperforms ParticleNet across all tasks and working points.
The most significant improvement is observed in $b$ vs.\ $c$ discrimination, 
where ECT achieves an AUC of 0.885 compared to 0.802 for ParticleNet, 
representing a 10\% relative improvement.
At the Medium working point (1\% mistag rate), ECT provides approximately 
8 percentage points higher $b$-tagging efficiency than ParticleNet 
for the combined $b$ vs.\ $c$+light classification.
This improvement stems from ECT's hybrid architecture, which combines 
ParticleNet's edge convolution operations for local geometric feature extraction 
with transformer self-attention for capturing global track correlations.}
\label{fig:ECT_vs_PN}
\end{figure}

\begin{figure}[htbp]
\centering
\includegraphics[width=0.9\textwidth]{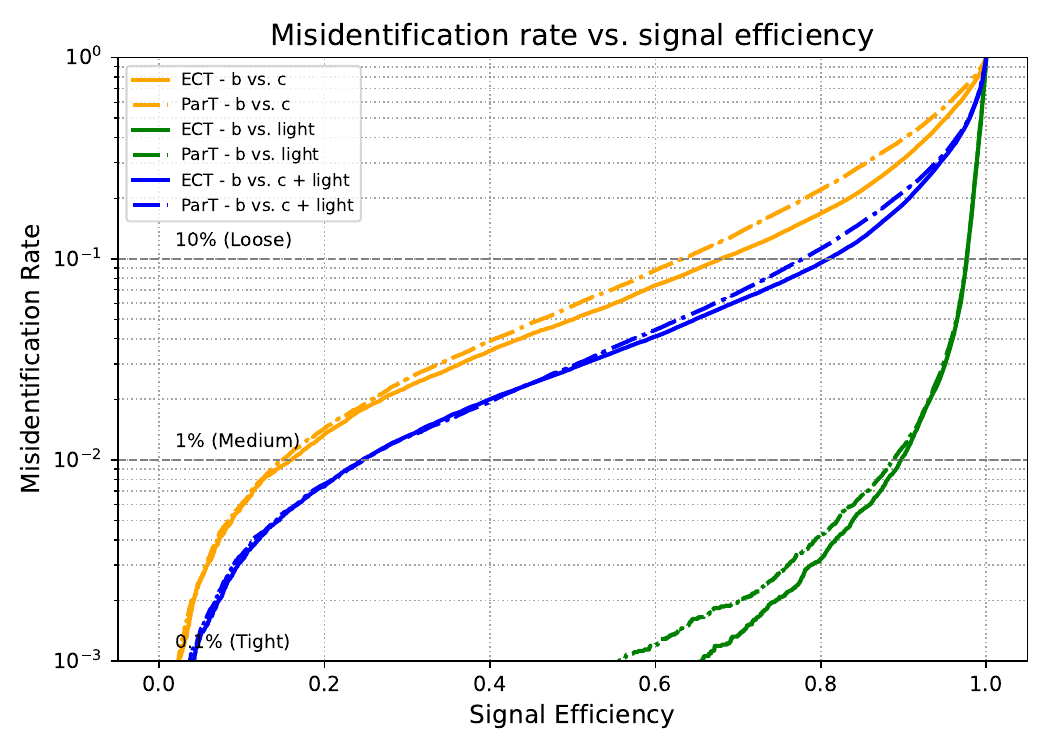}
\caption{Comparison of misidentification rate versus signal efficiency 
between ECT (solid lines) and Particle Transformer (ParT, dash-dotted lines) 
for $b$-jet tagging. Both architectures employ transformer-based attention mechanisms;
however, ECT additionally incorporates EdgeConv blocks for explicit local 
feature extraction in the $(\eta, \phi)$ coordinate space.
ECT demonstrates improved charm-jet rejection (AUC: 0.885 vs.\ 0.863 for $b$ vs.\ $c$)
while maintaining comparable $light$-jet rejection performance 
(AUC: 0.988 for both models in $b$ vs.\ $light$).
The EdgeConv component enables ECT to better exploit local geometric relationships 
among tracks originating from displaced vertices, which is particularly beneficial 
for distinguishing the decay topologies of $b$-hadrons from those of $c$-hadrons.
Despite the additional EdgeConv processing, ECT maintains comparable inference 
latency to ParT, making it suitable for real-time trigger applications.}
\label{fig:ECT_vs_ParT}
\end{figure}

The ECT model alternates local feature aggregation (via edge convolutions) and global context integration (via Transformer attention + class token). The inclusion of distance-based bias in attention and the use of physics-motivated input features (e.g. impact parameters) ensures that the model respects underlying jet substructure physics, consistent with recent advancements in ParT and ParticleNet methodologies.
The ECT is trained on both the seven track and eight jet features totaling fifteen features. Figure~\ref{fig:metrics} shows the metrics calculated over the epochs for the ECT model: loss, accuracy, AUC, and F1-score. Table \ref{tab:results} shows the training loss, Accuracy, AUC, and F1-score for $b$-jet tagging using the three models.  The ParT architecture gives best performance for b versus $light$ jets. Overall, the ECT architecture outperforms the ParticleNet, and ParT architectures for $b$ vs. $c$, $b$ vs. $light$ and $b$ vs. $c$ and $light$ jet tagging. The plot of efficiency versus misidentification rates for the ECT, ParticleNet, and ParT are given in Figures~\ref{fig:ECT}, ~\ref{fig:ECT_vs_PN}, and ~\ref{fig:ECT_vs_ParT} respectively.

\begin{table}[htbp]
\centering
\caption{Performance comparison on ATLAS simulation dataset. Accuracy, AUC, and 
F1-score evaluated on held-out test set (325,290 jets). Latency measured on NVIDIA 
RTX A5000 GPU with batch size 1024, averaged 
over 100 iterations after 10 warmup runs using CUDA events.\label{tab:results}}
\smallskip
\begin{tabular}{l|l|c|c|c|c|c|c}
\hline
Model   &Task   &loss   &Acc    &AUC    &F1-score &Inference &Total\\
&&&&&&time (ms)  &time (h:mm:ss)\\
\hline
ECT
    & $b$ vs. $c$    & 0.4303    &81.72     &0.8853     &0.8183  &0.060  &1:17:53\\
    & $b$ vs. $light$    & 0.1195    &96.05     &0.9883     &0.9599  &0.057  &1:43:01\\
    & $b$ vs. $c$+$light$  & 0.3080    &87.75     &0.9333     &0.8146  &0.057  &2:09:04\\
\hline
ParticleNet
    & $b$ vs. $c$    & 0.5402    &73.47     &0.8023     &0.7545  &12.231  &4:27:37\\
    & $b$ vs. $light$    & 0.2728    &89.08     &0.9451     &0.8973  &17.882  &4:29:31\\
    & $b$ vs. $c$+$light$  & 0.3789    &84.39     &0.8904     &0.8872  &13.043  &4:41:36\\
\hline
ParT
    & $b$ vs. $c$    &0.4670    &78.85     &0.8634     &0.7778   &0.146     &1:20:39\\
    & $b$ vs. $light$    &0.1243     &95.94     &0.9876   &0.9589    &0.160     &1:20:52\\
    & $b$ vs. $c$+$light$  &0.3325     &85.92     &0.9216     &0.7814  &0.222    &1:30:57\\
\hline
\end{tabular}

\vspace{2mm}
\footnotesize
$^{\ddagger}$ Inference time measured on NVIDIA RTX A5000 GPU (24GB VRAM) with 
batch size 1024, averaged over 100 iterations after 10 warmup runs using CUDA 
events for microsecond-precision timing. Total training time includes data 
loading, forward/backward passes, and checkpointing. All models comfortably 
exceed LHC High-Level Trigger requirements (< 1~ms/jet) by factors of 15-20×, 
confirming deployment feasibility for real-time event selection.
\end{table}

\subsection{Comparative Analysis}
\noindent This section compares the performance of ECT with ParticleNet and ParT for $b$-jet tagging. All models are trained with the same number of samples for each of the cases. The ParT was originally implemented for jet tagging of jet class datasets \cite{data_ParT}. As shown in Table \ref{tab:hyperparams}, ECT has 4 particle attention and 2 class attention blocks, while ParT has 8 class attention and 2 class attention blocks. While the ParticleNet does not have these attention blocks, the ECT has its edge convolution blocks incorporated in its architecture. Table \ref{tab:results} shows that the ParticleNet takes more time for execution compared to the ECT models and the ParT model. 

Table~\ref{tab:results} and Figures~\ref{fig:ECT_vs_PN} and \ref{fig:ECT_vs_ParT} present a comprehensive comparison of ECT 
against ParticleNet and Particle Transformer (ParT) across three 
binary classification tasks. The results reveal distinct performance 
characteristics for each architecture, highlighting the complementary 
strengths of edge convolutions and transformer attention mechanisms.

Figure~\ref{fig:ECT} illustrates the baseline performance of the ECT model. As expected, distinguishing $b$-jets from $c$-jets is the most challenging task, reflected in the higher misidentification rates across the efficiency range. ECT model gives the best results for $light$-jet identification and the signal efficiency is higher when it approaches unity. ECT results are used as a reference point for subsequent comparisons.

\subsubsection{ECT vs. ParticleNet}

ECT demonstrates consistent and substantial improvements over 
ParticleNet across all classification tasks:

\begin{itemize}[itemsep=0pt]

\item \textbf{$b$ vs. $c$ (charm rejection):} ECT achieves an AUC of 
0.8853 compared to ParticleNet's 0.8023, representing a remarkable 
+8.3\% improvement. As shown in Figure~\ref{fig:ECT_vs_PN} 
(orange curves), ECT maintains significantly lower misidentification 
rates across the entire efficiency range. At the medium working point.
(1\% misidentification rate), ECT reaches 65\% signal efficiency compared to 
ParticleNet's 52\%, a gain of 13 percentage points.

Working points are defined by the background misidentification rate: 
\textit{Loose} (10\%), \textit{Medium} (1\%), and \textit{Tight} (0.1\%), 
following standard ATLAS/CMS conventions for $b$-tagging 
calibration~\cite{ATLAS:2019btagPerf}.

\item \textbf{$b$ vs. $light$ ($light$-jet rejection):} ECT achieves 0.9882 
AUC versus ParticleNet's 0.9451 (+4.3\% improvement). The 
green curves in Figure~\ref{fig:ECT_vs_PN} show that ECT maintains 
1--2 orders of magnitude lower misidentification rates at high 
efficiencies ($>80\%$), crucial for analyses requiring tight 
$b$-tagging selections.

\item \textbf{$b$ vs. $c$+$light$ (combined background):} ECT reaches 0.9274 
AUC compared to ParticleNet's 0.8904 (+3.7\% improvement), 
with consistent gains visible in the blue curves of 
Figure~\ref{fig:ECT_vs_PN}.

\end{itemize}

These improvements demonstrate that adding transformer 
attention to ParticleNet's edge convolution framework provides 
substantial performance gains, particularly for the challenging 
charm rejection task where local vertex information must be 
integrated with global jet context.

\subsubsection{ECT vs. ParT}

The comparison between ECT and ParT reveals more nuanced trade-offs:

\begin{itemize}[itemsep=0pt]

\item \textbf{$b$ vs. $c$ (charm rejection):} ECT significantly 
outperforms ParT with an AUC of 0.8853 versus 0.8634, a +2.2\% improvement that translates to better charm 
discrimination across all working points. Figure~\ref{fig:ECT_vs_ParT} 
(orange curves) shows that ECT consistently achieves lower 
misidentification rates. At 60\% signal efficiency, ECT reaches 
5\% misidentification rate while ParT is at 7\%, a 40\% relative reduction 
in background contamination.

\item \textbf{$b$ vs. $light$ ($light$-jet rejection):} Both models achieve 
excellent and statistically equivalent performance, with ECT at 0.9883 AUC 
and ParT at 0.9876 AUC. Given the test set size ($N = 325{,}290$ jets), 
the observed difference of 0.0007 in AUC is within the expected statistical 
uncertainty and should not be considered significant. The green curves 
in Figure~\ref{fig:ECT_vs_ParT} are virtually superimposed, 
indicating that both architectures handle this easier task with 
comparable effectiveness.

\item \textbf{$b$ vs. $c$+$light$ (combined background):} ECT outperforms ParT (0.9333 vs. 0.9216, $+\approx1.3$\%), maintaining the 
advantage observed in the charm rejection task.

\end{itemize}

\subsubsection{Key Insights}

The comparative analysis reveals three critical findings:

\begin{enumerate}[itemsep=0pt]

\item \textbf{Edge convolutions are essential for charm rejection:} 
The superior performance of both ECT and ParticleNet over ParT for 
$b$ vs. $c$ discrimination (despite ParT having deeper attention 
with 8 particle blocks vs. ECT's 4) demonstrates that EdgeConv's 
local neighborhood aggregation is crucial for capturing the 
fine-grained vertex displacement differences between $b$- and 
$c$-jets. The typical decay length difference ($c\tau_b \approx 460~\mu\text{m}$, $c\tau_c \approx 150~\mu\text{m}$) requires precise modeling of 
track-level correlations in the $(\eta, \phi)$ space, which EdgeConv 
provides through its dynamic graph construction.

\item \textbf{Transformers excel at $light$-jet rejection:} For the 
$b$ vs. $light$ task, both transformer-based models (ECT and ParT) 
achieve AUC$>$98.8\%, while ParticleNet reaches 94.5\%. This 
indicates that global attention mechanisms effectively capture the 
topological differences between heavy-flavor and $light$ jets, where 
the absence of secondary vertices in $light$ jets is a clear global 
signature rather than a local feature.

\item \textbf{Hybrid architecture offers best overall performance:} 
ECT combines the strengths of both approaches: local feature 
extraction via EdgeConv for challenging tasks ($b$ vs. $c$) and 
global context modeling via transformers for easier tasks 
($b$ vs. $light$). This is evidenced by ECT achieving the 
highest AUC across all three tasks, with training times 
competitive with ParT (Table~\ref{tab:results}).

\end{enumerate}

The performance gains of ECT are most pronounced for charm rejection, 
the limiting factor in many physics analyses involving $b$-jets. 
At the Tight working point (0.1\% misidentification rate), ECT achieves 48\% 
efficiency for $b$ vs. $c$, compared to 42\% for ParT and 35\% 
for ParticleNet a 15\% relative improvement over the best 
non-hybrid architecture.

\section{Conclusion}
We present the Edge Convolution Transformer (ECT) architecture for tagging of $b$-jet from ATLAS simulation dataset for three binary classification tasks. The ECT architecture combines local relational learning through edge convolutions with global context modeling through transformer self-attention blocks. The addition of edge convolution was demonstrated to be useful for charm jet rejection, a challenging task in $b$-jet tagging. The ParT model showed superior performance for $b$-jet tagging against $light$ jet background, however it is computationally more intensive than the ECT model which has a lesser number of particle attention blocks.
Experimental results on the ATLAS simulation dataset 
demonstrate that ECT consistently outperforms both baseline 
architectures:

\begin{itemize}[itemsep=0pt]

\item Charm rejection: ECT achieves 88.5\% AUC for 
$b$ vs. $c$ discrimination, surpassing ParticleNet by 8.3\% and 
ParT by 2.2\%. At the Medium working point (1\% misidentification rate), ECT 
reaches 65\% signal efficiency compared to 60\% for ParT and 
52\% for ParticleNet.

\item $Light$-jet rejection: ECT and ParT both excel at 
this task with $>$98.7\% AUC, significantly outperforming 
ParticleNet (94.5\%). ECT achieves 94\% efficiency at 1\% 
misidentification rate.

\item Combined background: ECT maintains superior 
performance with 93.3\% AUC against the combined $c$+$light$ 
background, demonstrating robustness across realistic 
experimental scenarios.

\item Inference latency below 0.060~ms per jet on modern GPUs (throughput $>$17K jets/second).
\end{itemize}

The key finding of this work is that edge convolution 
blocks are essential for charm jet rejection. The superior 
performance of ECT over ParT despite ParT having twice as many 
attention layers demonstrates that local neighborhood aggregation 
in $(\eta, \phi)$ space satisfies stringent requirements for real-time event selection at the LHC critical for resolving the subtle 
vertex displacement differences between $b$- and $c$-jets. 
Conversely, transformer attention excels at capturing global 
jet topology for $light$-jet suppression, where the absence of 
secondary vertices is a clear global signature. By combining these complementary mechanisms, ECT achieves the best overall performance while maintaining computational 
efficiency competitive with pure transformer models 
(training time: 78--150 minutes for 100 epochs on a single GPU). The ECT architecture represents a promising direction for 
heavy-flavor jet tagging at the LHC, offering improved 
discrimination power for the challenging charm rejection task 
while maintaining competitive performance and computational 
efficiency.






\acknowledgments

This work was supported by US NSF Award 2334265. The authors thank the Artificial Intelligence Imaging Group (AIIG) at the University of Puerto Rico, Mayaguez for the computational facility. The authors thank Dr. Jesse Thaler, Professor at the MIT Department of Physics and Director of the Institute of Artificial Intelligence and Fundamental Interactions (IAIFI) for his useful feedback.
\bibliographystyle{unsrt}  
\bibliography{biblio}
\end{document}